\documentclass[pre,floatfix,preprint,superscriptaddress,showpacs]{revtex4}
\usepackage{graphicx}% Include figure files
\usepackage[english]{babel}
\usepackage[usenames]{color}
\usepackage{amsmath,amsfonts,amssymb,float}
\usepackage{dcolumn}% Align table columns on decimal point
\usepackage{bm}% bold math
\def\clf{Central Laser Facility, STFC Rutherford Appleton Laboratory, Didcot, OX11 0QX, United Kingdom}

\def\strathclyde{SUPA, Department of Physics, University of Strathclyde, Glasgow, G4 0NG, United Kingdom}
\def\ist{GoLP/Instituto de Plasmas e Fus\~ao Nuclear - Laboratorio Associado, Instituto Superior T\'ecnico, 1049-001 Lisbon, Portugal}
\def\llnl{Lawrence Livermore National Laboratory, Livermore, California, USA}
\def\standrews{University of St Andrews, St Andrews, Fife KY16 9AJ, United Kingdom}
\def\oxford{Department of Physics, University of Oxford, Oxford OX1 3PU, UK}
\def\dcti{DCTI/ISCTE Lisbon University Institute, 1649-026 Lisbon, Portugal}

\begin{document}

\title{Boosting the performance of Brillouin amplification at
  sub-quarter-critical densities via reduction of parasitic Raman
  scattering}

\author{R.M.G.M. Trines}
\affiliation{\clf}
\author{E.P. Alves}
\affiliation{\ist}
\author{K.A. Humphrey}
\affiliation{\strathclyde}
\author{R. Bingham}
\affiliation{\clf}
\affiliation{\strathclyde}
\author{R.A. Cairns}
\affiliation{\standrews}
\author{F. Fi\'uza}
\affiliation{\llnl}
\author{R.A. Fonseca}
\affiliation{\ist}
\affiliation{\dcti}
\author{L.O. Silva}
\affiliation{\ist}
\author{P.A. Norreys}
\affiliation{\oxford}
\affiliation{\clf}
\date\today

\begin{abstract}
Raman and Brillouin amplification of laser pulses in plasma have been
shown to produce picosecond pulses of petawatt power. In previous
tudies, filamentation of the probe pulse has been identified as the
biggest threat to the amplification process, especially for Brillouin
amplification, which employs the highest plasma densities. Therefore
it has been proposed to perform Brillouin scattering at densities
below $n_{cr}/4$ to reduce the influence of filamentation. However,
parastic Raman scattering can become a problem at such densities,
contrary to densities above $n_{cr}/4$, where it is suppressed. In
this paper, we investigate the influence of parasitic Raman scattering
on Brillouin amplification at densities below $n_{cr}/4$. We expose
the specific problems posed by both Raman backward and forward
scattering, and how both types of scattering can be mitigated, leading
to an increased performance of the Brillouin amplification process.
\end{abstract}
\maketitle

\section{Introduction}

Amplification of laser beams via parametric instabilities in plasma
(Raman and Brillouin scattering) has been proposed a number of times
\cite{maier66,milroy77,milroy79, capjack82,andreev89}, but came into
its own only relatively recently \cite{shvets99,kirkwood99,ping04,
  weber1,ren07,ping09,lancia,trines10,kirkwood11,trines11,toroker12}.
Brillouin scattering has also been used to transfer energy via the
Cross-Beam Energy Transfer scheme at the National Ignition Facility
\cite{kruer96,williams04,michel09,glenzer10,michel10,hinkel11,moody12}.
Both Raman and Brillouin scattering have been studied extensively in
the context of Inertial Confinement Fusion \cite{tanaka82,walsh84,
  vill87,mori94,langdon02,lindl04,hinkel05,froula07,michel11,glenzer11};
Raman scattering also in the context of wakefield acceleration
\cite{forslund85,decker94,rousseaux95,karl95,tzeng96,decker96,moore97,
  tzeng98,gordon98,matsuoka10}. Raman and Brillouin scattering are
processes where two electromagnetic waves at slightly different
frequencies propagating in plasma exchange energy via a plasma
wave. For Raman scattering, this is a fast electron plasma wave, while
for Brillouin scattering it is a slower ion-acoustic wave
\cite{forslund}. When it comes to laser beam amplification, Raman and
Brillouin scattering have different properties and serve different
purposes. Raman amplification yields the shortest output pulses and
the highest amplification ratios, but it is sensitive to fluctuations
in the experimental parameters and requires high accuracy in the
matching of laser and plasma frequencies. Brillouin amplification
yields lower peak intensities or amplification ratios, but is far more
robust to parameter fluctuations or frequency mismatch, more efficient
(as less laser energy stays behind in the plasma wave) and more
suitable for the production of pulses with a high total power or
energy.

Previous investigations into Raman and Brillouin scattering identified
filamentation as the most important limiting factor for succesful
amplification \cite{trines10,alves14}. This is especially true for
Brillouin amplification, since it employs higher plasma densities than
Raman amplification. Thus, it has been proposed to reduce the plasma
density for Brillouin amplification from $n_0/n_{cr} = 0.3$
\cite{weber1} to $n_0/n_{cr} = 0.05$ \cite{weber13,riconda13}, where
$n_0$ denotes the background plasma electron density and $n_{cr}$
denotes the critical density for the wave length of the pump laser.
However, while stimulated Raman scattering (SRS) is suppressed for
$n_0/n_{cr} = 0.3$, it is possible at any density below $n_0/n_{cr} =
0.25$ \cite{forslund}, and can be expected to interfere with the
Brillouin amplification process. Examples of strong longitudinal pulse
envelope modulations and intense prepulses preceding the amplified
probe pulse, all induced by Raman forward scattering, have been
observed before \cite{riconda13,alves14}. Therefore, we need to
investigate the influence of stimulated Raman scattering (both
backward and forward) on Brillouin amplification at
sub-quarter-critical densities. This will be done as follows. First,
we will give a summary of the self-similar theory of Brillouin
amplification in the strong-coupling regime
\cite{weber1,alves14}. Next, we will thoroughly analyse the results of
a particle-in-cell (PIC) simulation of a scenario where strong SRS is
likely to occur: laser beam intensities of $10^{16}$ W cm$^{-2}$ for a
waven length of 1 $\mu$m and $n_0/n_{cr} = 0.05$. Finally, we will
carry out a thorough parameter scan to identify the parameters for the
pump laser and the plasma column where the best results (highest
amplification factor, lowest relative level of parasitic SRS) can be
obtained. We will also investigate and discuss the impact of using
non-constant plasma density profiles, as proposed by Riconda \emph{et
  al.}  \cite{riconda13}.

\section{Self-similar theory of Brillouin amplification}

We start from a homogeneous plasma with electron number density $n_0$,
plasma frequency $\omega_p^2 = e^2 n_0/(\varepsilon_0 m_e)$, ion
plasma frequency $\omega_{pi} = \omega_p\sqrt{Z^2 m_e/m_i}$,
electron/ion temperatures $T_e$ and $T_i$, electron thermal speed
$v_T^2 = k_B T_e/m_e$, Debye length $\lambda_D = v_T/\omega_p$, and a
pump laser pulse with wave length $\lambda$, intensity $I$, frequency
$\omega_0 = 2\pi c/\lambda$, dimensionless amplitude $a_0 \equiv
8.55\times 10^{-10} \sqrt{g} \sqrt{I \lambda^2 [\mathrm{Wcm}^{-2}\mu
    \mathrm{m}^2]}$, where $g=1$ ($g=1/2$) denotes linear (circular)
polarisation, and wave group speed $v_g/c = \sqrt{1-\omega_p^2/
  \omega_0^2} = \sqrt{1-n_0/n_{cr}}$. Let the durations of pump and
probe pulse be given by $\tau_{pu}$ and $\tau_{pr}$, and define
$\gamma_B = (\sqrt{3}/2) [a_0 (v_g/c) \omega_{pi} \sqrt\omega_0
]^{2/3}$, the Brillouin scattering growth rate in the strong-coupling
regime \cite{forslund}. Expansion of the self-similar
coordinate $\xi$ of Ref. \cite{weber1}, and application of the energy
balance $a^2_{pr} \tau_{pr} = \eta a_0^2 \tau_{pu}$ yields:
\begin{align}
\label{eq:ssbril}
a_0(v_g/c) \omega_{pi} \tau_{pr} \sqrt{\omega_0 \tau_{pu}} &=
\sqrt{2g/\eta} \xi_B,\\
\label{eq:topt}
a_{pr}^2 \tau_{pr}^3 &= 2g\xi_B^2 [\omega_{pi}^2 \omega_0
  (1-\omega_{pe}^2/\omega_0^2)]^{-1},
\end{align}
where $\xi_B \approx 3.5$ is a numerical constant and $\eta$ denotes
the pump depletion efficiency. The physical interpretation of
Eq. (\ref{eq:ssbril}) is that the duration of the probe pulse is
similar to the time it takes the probe to deplete the
counterpropagating pump: for increasing pump intensity or probe
amplification (i.e. longer $\tau_{pu}$), pump depletion is more rapid
and $\tau_{pr}$ decreases. This allows one to tune the final probe
duration via the properties of the pump beam, similar to Raman
amplification \cite{trines11}. Eq. (\ref{eq:topt}) implies that the
initial probe pulse duration is not a free parameter: this equation
dictates the optimal initial probe pulse duration $\tau_{opt}$ for a
given initial probe pulse amplitude $a_1$.

From previous numerical work on Raman \cite{kim03,trines11} and
Brillouin amplification \cite{lehmann13}, it follows that if the probe
pulse is too short for its amplitude initially, it will reshape itself
first to fulfil Eq. (\ref{eq:topt}), and only then start to
amplify. Ensuring that the probe pulse fulfils Eq. (\ref{eq:topt})
from the start will speed up the amplification process and increase
the efficiency. For that reason, we will vary the plasma density, pump
intensity and the interaction length in our simulations, but the
initial probe pulse intensity will be chosen equal to the pump
intensity, and the probe duration will be treated as a dependent
parameter and calculated using Eq. (\ref{eq:topt}).

\section{Parasitic instabilities at sub-quarter-critical densities}

\subsection{Theory}

\begin{figure*}[ht]
\includegraphics[width=0.45\textwidth]{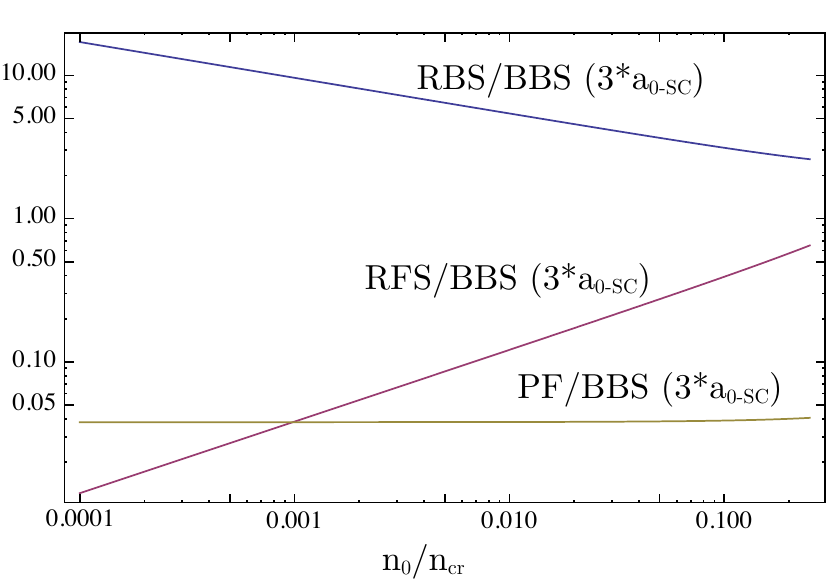}
\caption{Growth rates for Raman backward scattering (RBS), Raman
  forward scattering (RFS) and ponderomotive filamentation (PF),
  relative to the strong-couping Brillouin backward scattering growth
  rate, versus plasma density. The pump laser amplitude is determined
  by the threshold for strong coupling Brillouin scattering \cite{forslund}.}
\label{fig:1}
\end{figure*}

In order to assess the relative importance of various competing
processes, we have calculated the growth rates for Raman backward
scattering ($\gamma_{RBS} = (a_0/2)\sqrt{\omega_0\omega_{pe}}$), Raman
forward scattering ($\gamma_{RFS} = (a_0/2\sqrt{2})\omega_{pe}^2/
\omega_0$), ponderomotive filamentation ($\gamma_{PF} = (5/4) a_0
\omega_{pi}$ in the limit $\gamma_{PF} \gg k_{PF} v_T \sqrt{m_e/m_i}$,
\cite{kaw73}), relativistic filamentation ($\gamma_{RF} = (a_0^2/8)
\omega_{pi}^2/\omega_0$, \cite{max74}) and strong-coupling Brillouin
backward scattering ($\gamma_B$, see above), for $0.0001 \leq
n_0/n_{cr} \leq 0.1$. The pump field amplitude was chosen to be three
times the threshold value for strong-coupling Brillouin scattering
\cite{forslund}, i.e. $a_0^2 = 36(v_T/c)^3 (n_{cr}/n_0)
\sqrt{1-n_0/n_{cr}} \sqrt{Zm_e/m_i}$. We made this particular choice
because we found earlier that the pump pulse amplitude should be as
low as possible, but still above the strong-coupling threshold, for
optimal results \cite{alves14}. With these adjustments to the pump
intensity, the density dependence of the various growth rates is as
follows:

\begin{align}
\frac{\gamma_{BBS}}{\omega_0} &= \sqrt{3} \frac{v_T}{c} \left(
\frac{m_e}{m_i} \right)^{1/2} \left(2 Z^2 \right)^{1/3} = \mathrm{const.},\\
\frac{\gamma_{RBS}}{\omega_0} &=2\sqrt{2} \left( \frac{m_e}{m_i} \right)^{1/4}
\left( \frac{v_T}{c} \right)^{3/2} \left( \frac{\omega_0}{\omega_{pe}} 
\right)^{1/2} \propto \left( \frac{n_0}{n_{cr}} \right)^{-1/4},\\
\frac{\gamma_{RFS}}{\omega_0} &= 2\left( \frac{m_e}{m_i} \right)^{1/4}
\left( \frac{v_T}{c} \right)^{3/2} \frac{\omega_{pe}}{\omega_0}
\propto \left( \frac{n_0}{n_{cr}} \right)^{1/2},\\
\frac{\gamma_{PF}}{\omega_0} &= 5\left( \frac{m_e}{m_i} \right)^{3/4}
\left( \frac{v_T}{c} \right)^{3/2} = \mathrm{const.}, \\
\frac{\gamma_{RF}}{\omega_0} &= 2\left( \frac{m_e}{m_i} \right)^{3/2}
\left( \frac{v_T}{c} \right)^3 = \mathrm{const.}
\end{align}

These growth rates are plotted in Figure \ref{fig:1}; all growth rates
are shown relative to the Brillouin scattering growth rate. For our
particular configuration, we find that the growth rate for the
ponderomotive filamentation does not change with density, while the
RBS growth rate increases and the RFS growth rate decreases with
decreasing plasma density. As will be confirmed in our simulation
results below, a density of $n_0/n_{cr} = 0.05$ is really too high,
driving too much RFS, while much better results can be obtained for
densities around $n_0/n_{cr} = 0.01$. At even lower densities,
e.g. $n_0/n_{cr} = 0.001$, one has to worry that the growth rate for
Brillouin scattering becomes too low for this process to be useful,
while the plasma frequency becomes low enough that the (anti-)Stokes
side bands of Raman scattering, located at $\omega_0 \pm \omega_{pe}$,
may fall within the bandwidth of the probe pulse and may be directly
driven by it. This fixes the useful density interval to roughly $0.005
\leq n_0/n_{cr} \leq 0.02$.

It should be noted that the adjusted growth rate of Raman backscatter
increases for decreasing plasma density because the pump laser
intensity is increased in order to remain above the threshold for
strong-coupling Brillouin scattering. However, we do not observe a
corresponding increase in the overall level of RBS in our numerical
simulations. It is conjectured that RBS saturates at lower densities
due to wave breaking of the RBS Langmuir wave, since the amplitude threshold
for wave breaking scales as $\sqrt{n_0/n_{cr}}$.

\subsection{Numerical simulations}

We have carried out a sequence of one-dimensional particle-in-cell
(PIC) simulations using the code OSIRIS
\cite{osiris1,osiris2,osiris3}. Parameters varied in these simulations
are the plasma density ($n_e/n_{cr} = 0.05$ or 0.01), the pump
intensity ($I_0 = 10^{16}$, $10^{15}$ or $10^{14}$ W cm$^{-2}$) and
the interaction length. The initial seed pulse intensity was chosen to
be the same as the pump intensity, and the seed duration was be half
the value predicted by (\ref{eq:topt}). The plasma column was given a
constant density, and had a fixed length. The simulations were
conducted in a static window, since we needed to study the pump pulse
reflection due to Raman backward scattering. The computational demands
of the simulations forced us to conduct them in one dimension; even
so, useful trends could be unearthed. Although filamentation cannot
be modelled in 1-D simulations, we expect that it will decrease in
importance for lower densities, as its growth rate scales quite
quickly with density: $\gamma_f = (a_0^2/8)(\omega_p^2/\omega_0)$
\cite{kaw73,max74}. The plasma profile is basically a plateau with
length $L$ and very steep ramps; the pump laser pulse has an FWHM
duration of $L/c$. We perform each of these simulations for 3
different plasma plateau lengths corresponding to $10$, $30$ and $100$
RBS growth lengths for the pump pulse, where the growth length is
given by $L_{RBS} =c/\gamma_{RBS} = 2c/(a_0\sqrt{\omega_0
  \omega_p})$. This ensures that the levels of premature pump RBS are
comparable between various pulse intensities and plasma densities.  We
use a spatial resolution of $dx = \lambda_D/2$, in order to accurately
describe the thermal nature of the ion acoustic wave, and use $100$
particles per cell per species with cubic interpolation for the
current deposition. Absorbing boundary conditions are imposed for the
electromagnetic fields.

\begin{figure*}[ht]
\includegraphics[width=0.45\textwidth]{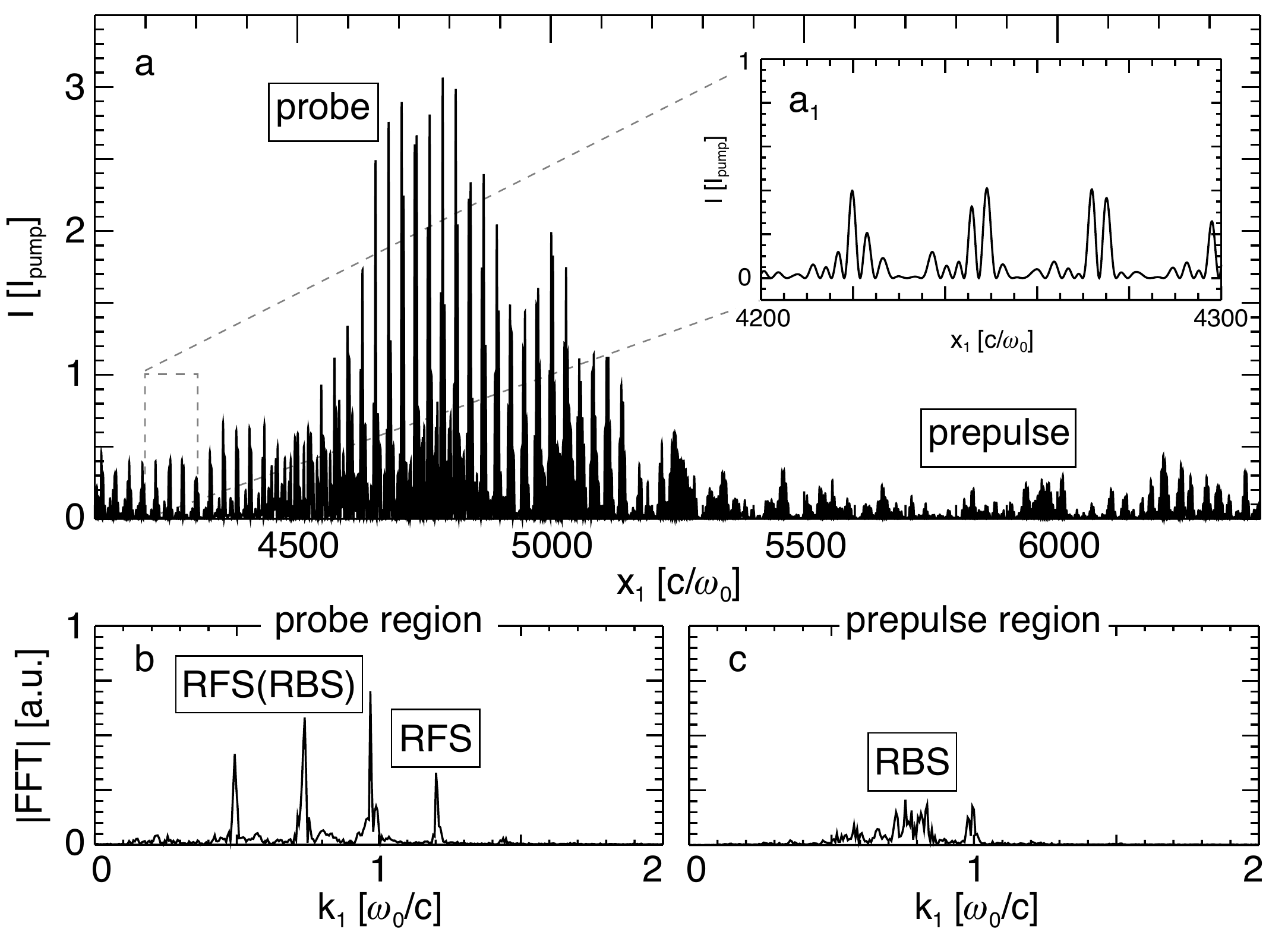}
\includegraphics[width=0.45\textwidth]{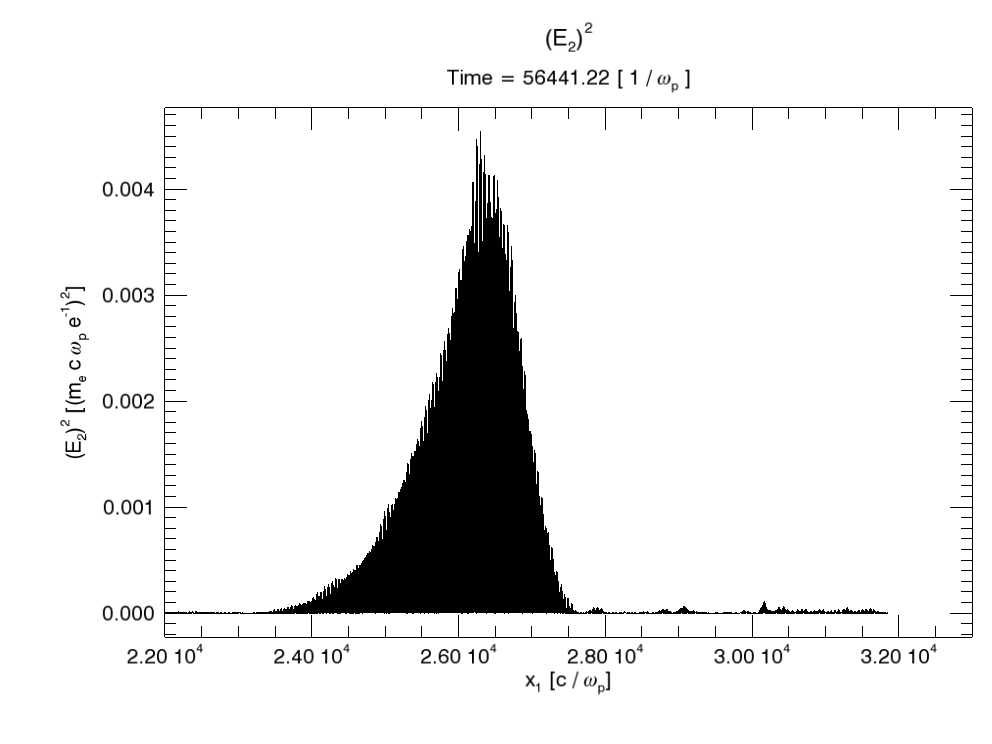}
\caption{Left: Parasitic stimulated Raman scattering occurring during
  Brillouin amplification in the sub-quarter-critical density regime
  ($n_0/n_{cr}=0.05$). Pulse intensities are $10^{16}$ W cm$^{-2}$,
  the pump pulse duration is 2.7 ps and the interaction length is 0.8
  mm. Pump-induced RBS/RFS and probe-induced RFS are shown in (a);
  inset a$_1$ reveals the development of incoherence at the probe
  tail. Frames (b) and (c) show the spectral signatures of the probe
  and prepulse regions, respectively. Right: amplified pulse
  corresponding to a simulation using a pump intensity of $10^{16}$ W
  cm$^{-2}$, $n_0/n_{cr}=0.01$ and 100 RBS growth lengths (5 mm
  interaction length), corresponding to the best result in Table
  \ref{table:1}. The reduction in parasitic RBS and the improved pulse
  quality are obvious, even after a much longer propagation through
  plasma.}
\label{fig:2}
\end{figure*}

To investigate the deleterious influence of Raman back- and forward
scattering (RBS and RFS), we have carried out an exploratory 1-D
static-window simulation at using typical parameters: a plasma slab
with electron density $n_0/n_{cr} = 0.05$ and length 0.8 mm and laser
pulses with intensities of $10^{16}$ W/cm$^2$ and a duration (pump) of
2.7 ps.  The results of this simulation are displayed in Figure
\ref{fig:2}, left. As shown in Figure \ref{fig:2}(a), RBS was found to
generate a large prepulse to the growing probe pulse, spoiling its
contrast, while RFS causes the probe pulse envelope to be strongly
modulated, rendering it about as dangerous as filamentation. A Fourier
analysis of the $k$-spectrum of the pulses, shown in
Fig. \ref{fig:2}(b) and (c), reveals that the pump pulse mostly
suffers from Raman backward scattering, while Raman forward scattering
is dominant in the probe pulse. A close inspection of all Raman
scattering occurring during Brillouin amplification found that the
growth of the probe pulse saturates due to high levels of Raman
forward scattering, rather than Raman backscattering. If the level of
RFS in the probe pulse becomes non-linear, the coherence of the probe
pulse's carrier wave, and thus the coupling between pump and probe, is
lost, and probe amplification stops; this can be seen in Figure
\ref{fig:2}(a1). It was also found that high levels of pump RFS are a
good indicator of non-linear probe RFS. Fortunately, reduction of RFS
can be achieved via a reduction in plasma density, which effects the
RFS growth rate more than any other growth rate: $\gamma_{RBS} \propto
a_0 \sqrt{\omega_0 \omega_p} \propto n_0^{1/4}$, $\gamma_B \propto
n_0^{1/3}$ while $\gamma_{RFS} \propto a_0 \omega_p^2 /\omega_0
\propto n_0$. It follows that lowering the plasma density even
further, e.g. to $n_0/n_{cr} = 0.01$, will immediately improve the
pump-to-probe amplification ratio and energy transfer.

From the results of our first simulation, it is obvious that
stimulated Raman scattering needs to be controlled, if one wishes to
conduct brillouin amplification at densities below $n_0/n_{cr} \leq
0.25$. To this end, we have conducted a sequence of 1-D
particle-in-cell simulations, where we varied the pump laser
intensity, the plasma density, and the interaction length, as
described above. A summary of all sub-quarter-critical simulations,
listing the pump-to-probe amplification ratio for each, is given in
Table \ref{table:1}. It is immediately clear that a plasma density of
$n_0/n_{cr} = 0.01$ yields better results than $n_0/n_{cr} = 0.05$;
this is found to be mainly caused by a reduction in RFS, which delays
saturation of the growing probe. An intensity of $10^{15}$ W cm$^{-2}$
also yields better results than an intensity of $10^{16}$ W cm$^{-2}$:
the final intensity may have a lower absolute value, but the relative
compression and amplification ratios are much higher. Results
deteriorate again for $10^{14}$ W cm$^{-2}$, but this is possibly
because these simulations could have been continued beyond 100 RBS
growth lengths. As it happens, RFS is the main limiting instability
rather than RBS, and the RFS growth length increases faster than the
RBS growth length when the density decreases, thus even an interaction
distance of 100 RBS growth lengths is ``too short'' for RFS to reach
problematic levels.

\begin{table}[ht]

\begin{tabular}{|c||c|c|c|c|c|c|}
\hline
 & \multicolumn{3}{|c|}{$0.05*n_{cr}$} &
  \multicolumn{3}{|c|}{$0.01*n_{cr}$} \\
\hline
 & 10 & 30 & 100 & 10 & 30 & 100 \\
\hline
%$10^{13}$ & 1.69w & 2.54w & 1.92w & w & w & w \\
%\hline
$10^{14}$ & 1.33 & 2.33 & 2.47 & 1.13w & 1.40w & 3.16w \\
\hline
$10^{15}$ & 1.35 & 2.06 & 4.76 & 1.09 & 1.46 & 6.3 \\
\hline
$10^{16}$ & 1.11 & 2.80 & 2.35 & 1.02 & 1.14 & 3.02 \\
\hline
\end{tabular}

\caption{Amplification ratio $I_f/I_0$ for plasma densities below
  $0.25*n_{cr}$ versus pump intensity, interaction length in terms of
  RBS e-foldings, and plasma density. Left column: pump intensity in
  W/cm$^2$. Top row: plasma density. Second row: interaction
  length. The suffix `w' implies that the interaction takes place in
  the weak-couping regime. The configurations of Refs. \cite{weber13,
    riconda13} correspond to $10^{16}$ W/cm$^2$, $0.05*n_{cr}$ and 65
  e-foldings.}
\label{table:1}
\end{table}

The results of Table \ref{table:1} can be summarised as follows. (i)
For an interaction length of $10*L_R$, parasitic RBS remains at a low
level, but the amplification ratio is very small; the highest ratio
obtained is 1.35 for $n_e/n_{cr} = 0.05$ and a $10^{15}$ W cm$^{-2}$
pump pulse. (ii) For an interaction length of $30*L_R$ or $100*L_R$,
the amplification ratio improves, e.g. a ratio of 6.3 was obtained for
$n_e/n_{cr} = 0.01$ and a $10^{15}$ W cm$^{-2}$ pump pulse, but the
level of RBS increases proportionally, reaching up to 10\% of the pump
energy for a $100*L_R$ interaction length. (iii) The amplification
ratio increases for decreasing intensity, but only if the pump
intensity/amplitude is above the threshold for strong coupling, given
by $a_0^2 > 4 (v_e^2/c^2) (\omega_0 k_0 c_s /\omega_{pe}^2)$, where
$v_e^2 = T_e/m_e$ and $c_s^2 = ZT_e/m_i$. For pump intensities below
this threshold, the amplification ratio decreases again. (iv)
Simulations at $n_e/n_{cr} = 0.15$ had to be abandoned since the level
of parasitic RBS became intolerable. For comparison, Riconda and Weber
\cite{weber13,riconda13} used $n_e/n_{cr} = 0.05$, a $10^{16}$ W
cm$^{-2}$ pump pulse and a $65*L_R$ interaction length, and obtained
amplification ratios of 8--10 (for a $10^{16}$ W cm$^{-2}$ initial
probe intensity), but at a cost of a high level of unwanted RBS,
almost equalling the level of (wanted) BBS in some cases, as well as
significant modulation of the seed envelope by RFS, as is clear from
the frequency spectra in Ref. \cite{riconda13}.

Raman forward scattering puts an upper limit on the compression and
amplification ratios that can be reached for sub-quarter-critical
densities. For example, it was shown that a probe pulse could be
amplified from $10^{16}$ W cm$^{-2}$ to $10^{17}$ W cm$^{-2}$ and
from $10^{17}$ W cm$^{-2}$ to $5\times 10^{17}$ W cm$^{-2}$ in two
separate simulations at $n_e/n_{cr} = 0.05$ \cite{weber13,riconda13}.
However, this does not imply that amplification from $10^{16}$ W cm$^{-2}$
to $5\times 10^{17}$ W cm$^{-2}$ is possible for the same probe pulse,
because the probe RFS generated during the first stage will saturate
the amplification during the second stage well before an intensity of
$5\times 10^{17}$ W cm$^{-2}$ is reached.

\begin{figure*}[ht]
\includegraphics[width=0.45\textwidth]{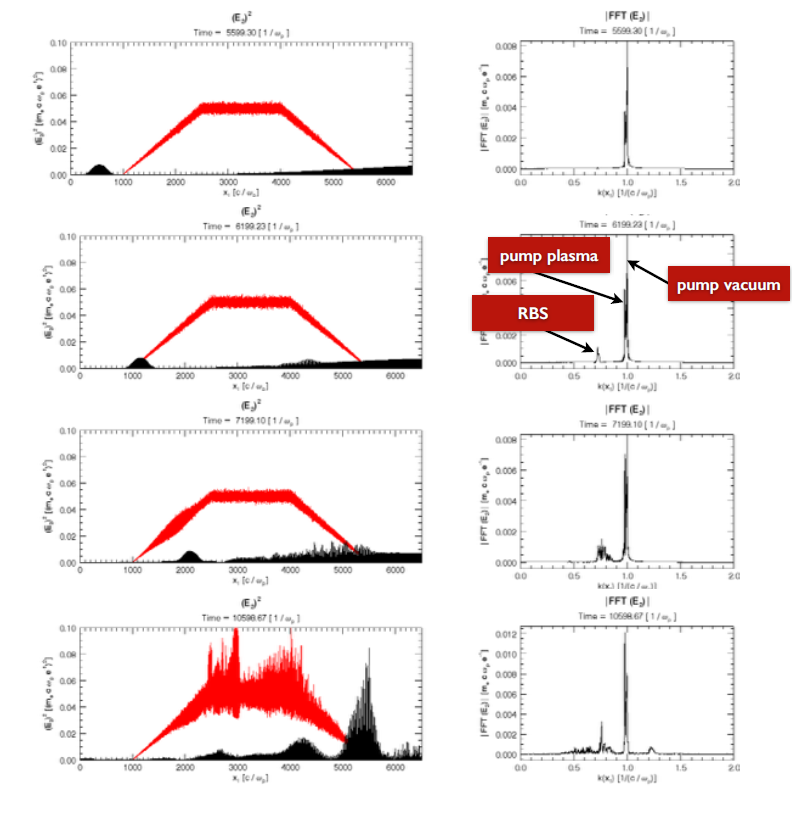}
\caption{Brillouin amplification using a ``trapezoid'' profile with a
  maximum density of $n_0/n_{cr} = 0.05$, a central plateau of 0.24 mm
  with ramps of 0.24 mm on either side. The pump pulse has an
  intensity of $10^{16}$ W cm$^{-2}$ and a duration of 5.3 ps. The
  pump-to-probe amplification ratio reaches a maximum of 10, which is
  higher than comparable scenarios with a constant plasma density
  profile. This improvement is mainly attributed to the reduction of
  RFS due to the presence of the density ramps.}
\label{fig:3}
\end{figure*}

%[Get some nice figures from Paulo showing plateau versus ramp.]

As shown above, lowering the plasma density will immediately improve
the pump-to-probe amplification ratio and energy transfer, but it may
also reduce the Brillouin backscattering growth rate, especially at
the beginning of the interaction when the probe intensity is still
low. Using a plasma density profiles with a ``ramp'' rather than a
``plateau'', with the highest plasma density facing the probe pulse,
as proposed by Weber, Riconda \emph{et al.}  \cite{weber13,riconda13},
could be a good compromise in this case. We investigated this in
a simulation with a trapezoidal density profile (a plateau of 0.24 mm
with ramps of 0.24 mm on either side) instead of a constant plasma
density throughout. A significant reduction in premature pump RFS was
found, causing an improvement in probe growth (amplification factors
of up to 10 were found for ``ramp'' profiles), simply because the
average plasma density is lower for a ``ramp'' profile than for a
``plateau''. Brillouin growth is still kickstarted by the high plasma
density at the end facing the incoming probe pulse. Results are shown
in Figure \ref{fig:3}. It should be noted that frequency matching
between pump, probe and ion-acoustic frequencies is not really an
issue here, even in the presence of a density ramp, since the
ion-acoustic frequency is so small that the frequency difference
between the pump and probe pulses is always fully covered by the
bandwidth of the probe pulse. Conversely, the non-constant plasma
density and electron plasma frequency may strongly impact the growth
of all forms of Raman scattering, which is rather beneficial in this
case.

\section{Conclusions}

We have studied Brillouin amplification of short laser pulses in
plasma at electron densities $n_0/n_{cr} < 0.25$. At such densities,
filamentation of the growing probe laser pulse is reduced compared to
e.g. $n_0/n_{cr} = 0.3$, but stimulated Raman scattering, which is
inhibited for $n_0/n_{cr} > 0.25$, suddenly becomes possible and
introduces extra complications. Raman backscattering of the pump pulse
adds a large pre-pulse to the amplified probe, while Raman forward
scattering of the probe itself causes strong envelope modulations and
a reduction of pulse quality. Even worse, non-linear Raman forward
scattering destroys the coherence of the probe pulse's carrier wave,
inhibiting further Brillouin amplification. Therefore, parasitic Raman
scattering needs to be reduced at all cost in order to boost Brillouin
amplification at sub-quarter-critical plasma densities.

Fortunately, the RFS growth rate scales much faster with the plasma
density than the BBS growth rate ($n_0$ versus $n_0^{1/3}$), so
reducing the plasma density will immediately reduce RFS levels without
compromising the Brillouin amplification process too much. We have
performed a range of 1-D particle-in-cell simulations where we varied
the pump laser intensity, the plasma density and the interaction
length. The simulation results showed that lowering either the plasma
density or the pump intensity led to a significant improvement in the
amplification and compression ratios, as well as the quality of the
amplified pulse. The best result obtained was for $n_0/n_{cr} = 0.01$
and a pump intensity of $10^{15}$ W cm$^{-2}$, although there are
strong indications that even better results can be obtained by
increasing the interaction length for the simulations at $10^{14}$ W
cm$^{-2}$ pump intensity and $n_0/n_{cr} = 0.01$. In particular, we
conclude that Brillouin amplification should be conducted at densities
for which RFS is either impossible ($n_0/n_{cr} > 0.25$) or
unimportant ($n_0/n_{cr} \leq 0.01$). For $0.01 < n_0/n_{cr} < 0.25$,
the disadvantage of increased pump RBS and probe RFS is more serious
than the advantage of reduced probe filamentation.

As a compromise between using a higher density to improve Brillouin
scattering and a lower density to reduce Raman scattering, one can use
a plasma density profiles with a ``ramp'' rather than a ``plateau'',
with the highest plasma density facing the probe pulse. This will
stimulate Brillouin scattering during the early stages of the
interaction, when the probe intensity is still low, while reducing
Raman forward scattering later on, when the probe intensity is much
higher. Initial simulations of this scenario showed a reduction in RFS
accompanied by an improvement in probe amplification and quality, so
the use of tailored plasma density profiles deserves further
investigation.

This work was supported by the STFC Central Laser Facility, the STFC
Centre for Fundamental physics and by EPSRC through grant
EP/G04239X/1. We would like to thank R. Kirkwood for stimulating
discussions and the OSIRIS consortium for the use of OSIRIS. We
acknowledge PRACE for providing access to the resource SuperMUC based
in Germany at the Leibniz research center. Simulations were performed
on the Scarf-Lexicon Cluster (STFC RAL), the IST Cluster (IST Lisbon)
and SuperMUC (Leibniz Supercomputing Centre, Garching, Germany).

\end{document}